\documentclass[12pt,epsf]{article}
\usepackage{graphicx,amsmath,amssymb}
\begin{document}

\def\a{\alpha}
\def\b{\beta}
\def\c{\varepsilon}
\def\d{\delta}
\def\e{\epsilon}
\def\f{\phi}
\def\g{\gamma}
\def\h{\theta}
\def\k{\kappa}
\def\l{\lambda}
\def\m{\mu}
\def\n{\nu}
\def\p{\psi}
\def\q{\partial}
\def\r{\rho}
\def\s{\sigma}
\def\t{\tau}
\def\u{\upsilon}
\def\v{\varphi}
\def\w{\omega}
\def\x{\xi}
\def\y{\eta}
\def\z{\zeta}
\def\D{\Delta}
\def\G{\Gamma}
\def\H{\Theta}
\def\L{\Lambda}
\def\F{\phi}
\def\P{\Psi}
\def\S{\Sigma}

\def\o{\over}
\def\beq{\begin{eqnarray}}
\def\eeq{\end{eqnarray}}
\newcommand{\gsim}{ \mathop{}_{\textstyle \sim}^{\textstyle >} }
\newcommand{\lsim}{ \mathop{}_{\textstyle \sim}^{\textstyle <} }
\newcommand{\vev}[1]{ \left\langle {#1} \right\rangle }
\newcommand{\bra}[1]{ \langle {#1} | }
\newcommand{\ket}[1]{ | {#1} \rangle }
\newcommand{\EV}{ {\rm eV} }
\newcommand{\KEV}{ {\rm keV} }
\newcommand{\MEV}{ {\rm MeV} }
\newcommand{\GEV}{ {\rm GeV} }
\newcommand{\TEV}{ {\rm TeV} }
\def\diag{\mathop{\rm diag}\nolimits}
\def\Spin{\mathop{\rm Spin}}
\def\SO{\mathop{\rm SO}}
\def\O{\mathop{\rm O}}
\def\SU{\mathop{\rm SU}}
\def\U{\mathop{\rm U}}
\def\Sp{\mathop{\rm Sp}}
\def\SL{\mathop{\rm SL}}
\def\tr{\mathop{\rm tr}}

\def\IJMP{Int.~J.~Mod.~Phys. }
\def\MPL{Mod.~Phys.~Lett. }
\def\NP{Nucl.~Phys. }
\def\PL{Phys.~Lett. }
\def\PR{Phys.~Rev. }
\def\PRL{Phys.~Rev.~Lett. }
\def\PTP{Prog.~Theor.~Phys. }
\def\ZP{Z.~Phys. }

\newcommand{\bea}{\begin{eqnarray}}   
\newcommand{\eea}{\end{eqnarray}}
\newcommand{\bear}{\begin{array}}  
\newcommand {\eear}{\end{array}}
\newcommand{\bef}{\begin{figure}}  
\newcommand {\eef}{\end{figure}}
\newcommand{\bec}{\begin{center}}  
\newcommand {\eec}{\end{center}}
\newcommand{\non}{\nonumber}  
\newcommand {\eqn}[1]{\beq {#1}\eeq}
\newcommand{\la}{\left\langle}  
\newcommand{\ra}{\right\rangle}
\newcommand{\ds}{\displaystyle}
\def\SEC#1{Sec.~\ref{#1}}
\def\FIG#1{Fig.~\ref{#1}}
\def\EQ#1{Eq.~(\ref{#1})}
\def\EQS#1{Eqs.~(\ref{#1})}
\def\TEV#1{10^{#1}{\rm\,TeV}}
\def\GEV#1{10^{#1}{\rm\,GeV}}
\def\MEV#1{10^{#1}{\rm\,MeV}}
\def\KEV#1{10^{#1}{\rm\,keV}}
\def\lrf#1#2{ \left(\frac{#1}{#2}\right)}
\def\lrfp#1#2#3{ \left(\frac{#1}{#2} \right)^{#3}}


\baselineskip 0.7cm

\begin{titlepage}

\begin{flushright}
TU-887\\
IPMU11-0127
\end{flushright}

\vskip 1.35cm
\begin{center}
{\large \bf 
Low-scale Supersymmetry from Inflation
}
\vskip 1.2cm
Kazunori Nakayama$^a$
and
Fuminobu Takahashi$^{b,c}$

\vskip 0.4cm

{\it $^a$Department of Physics, University of Tokyo, Tokyo 113-0033, Japan}\\
{\it $^b$Department of Physics, Tohoku University, Sendai 980-8578, Japan}\\
{\it $^c$Institute for the Physics and Mathematics of the universe,
University of Tokyo, Kashiwa 277-8568, Japan}\\

\vskip 1.5cm

\abstract{ We investigate an inflation model with the inflaton being
  identified with a Higgs boson responsible for the breaking of
  U(1)$_{\rm B-L}$ symmetry.  We show that supersymmetry must remain a
  good symmetry at scales one order of magnitude below the inflation
  scale, in order for the inflation model to solve the horizon and
  flatness problems, as well as to account for the observed density
  perturbation. The upper bound on the soft supersymmetry breaking
  mass lies between $1$\,TeV and $\TEV{3}$.  Interestingly, our
  finding opens up a possibility that universes with the low-scale
  supersymmetry are realized by the inflationary selection.  Our
  inflation model has rich implications; non-thermal leptogenesis
  naturally works, and the gravitino and  moduli problems as well as the
  moduli destabilization problem can be solved or ameliorated; the standard-model
  higgs boson receives a sizable radiative correction if the supersymmertry
  breaking takes a value on the high side $\sim \TEV{3}$.}
\end{center}
\end{titlepage}

\setcounter{page}{2}

\section{Introduction}

The inflationary paradigm~\cite{Guth:1980zm} has been well established
so far. A number of theoretical difficulties of the standard big bang
cosmology are naturally circumvented by the exponential expansion of
the universe during inflation, and more important, the quantum
fluctuation of the inflaton field can account for the observed density
perturbation.

Despite the success of the inflationary paradigm, it has been
considered extremely challenging to answer the question, what is the
inflaton. If the inflaton is just a gauge singlet with extremely weak
interactions with the standard-model particles, it would be almost
impossible to identify the inflaton in a laboratory experiment. One
way to avoid this conclusion is to build a successful inflation model
in the framework of the standard model (SM) or its extensions.  In the
SM, the Higgs boson $\phi_{\rm SM}$ is the only scalar field, and
therefore a candidate for the inflaton.  It is indeed possible to
build an inflation model using $\phi_{\rm SM}$, relying on a
non-canonical kinetic term~\cite{Nakayama:2010sk,Germani:2010gm} and/or a non-minimal
coupling to gravity~\cite{Bezrukov:2007ep,Ferrara:2010yw}.

Since the discovery of neutrino oscillations, the right-handed
neutrinos, $\nu_R$, are usually incorporated in the minimal extension
of the SM to explain the small, but non-vanishing neutrino masses.  In
particular, the extremely light neutrino mass scale can be naturally
accounted for by the see-saw mechanism~\cite{seesaw}, which requires
the heaviest right-handed neutrino at a scale of $\GEV{15}$ close to the GUT
scale. With the addition of the three right-handed neutrinos, it is
then reasonable to introduce the U(1)$_{\rm B-L}$ gauge symmetry which
is required by the charge quantization condition and is also motivated
by the GUT gauge group such as SO(10).  Thus, we consider the
framework, SM+$\nu_R$+U(1)$_{\rm B-L}$, as the minimal extension of
the SM. In this theoretical framework, we have another candidate for
the inflaton, namely the Higgs boson, $\phi_{\rm B-L}$, which is responsible
for the breaking of the U(1)$_{\rm B-L}$ symmetry.  In this paper we
explore a possibility that the Higgs boson $\phi_{\rm B-L}$ plays a
role of the inflaton and discuss its implications.

The SM has been  successful in explaining numerous experimental data
with a great accuracy, and there is no hard evidence for physics
beyond the SM (with neutrino masses included). On the other hand, it
has been known that there is a gauge hierarchy problem in the SM,
which was the motivation to consider the physics beyond SM such as
supersymmetry (SUSY). However, after the LEP experiment, the
supersymmetric extension of SM (SSM) turned out to be not free of
fine-tunings.  Indeed, typically a fine-tuning at the percent level is
required for the correct electroweak breaking, which casts doubt on
the conventional naturalness argument as the correct guiding principle
for understanding the physics at and beyond the weak scale.

On the other hand, it is now widely accepted that SUSY should appear
at a certain energy scale, which may be much higher than the weak
scale, because the string theory, the most qualified candidate for the
unified theory including gravity, requires supersymmetry for
theoretical consistency, and it may remain in the effective 4D theory
below the compactification scale.  Further, the recent observation of
a cosmological constant within the anthropic
window~\cite{Weinberg:1987dv} strongly suggests the presence of the
string landscape.  Motivated by these considerations, we do not rely
on the conventional naturalness argument for building an inflation
model. For instance we do not care much about the fine-tuning needed
to make the inflaton potential flat, because such a tuning may be
easily compensated by the subsequent exponential expansion of the
universe during inflation, and because clearly we cannot live in the
universe which does not experience inflation.  Instead, we take the
existence of the inflationary phase (driven by $\phi_{\rm B-L}$ in the
model considered below) as a guiding principle.  Also we assume the
presence of SUSY, but we leave the SUSY breaking scale as a free
parameter since it may be subject to the distribution of vacua in the
landscape or anthropic selection. Indeed as we will see, the SUSY
should remain a good symmetry at scales one order of magnitude below
the inflation scale for the inflation model to be
successful. Typically the soft SUSY breaking masses for the SSM
particles lie between $1$\,TeV and $\TEV{3}$, whose precise
value depends on the B$-$L breaking scale and the inflaton
potential. If there is a bias toward high-scale SUSY breaking in the landscape,
the SUSY breaking masses may be close to $\TEV{3}$.
It is intriguing that the low-scale SUSY emerges as a
result of the inflationary selection, irrespective of the gauge
hierarchy problem.\footnote{Here and in what follows, the low-scale SUSY means
that SUSY remains a good symmetry at scales much smaller than the Planck or string
scale.}

\vspace{5mm}

Before closing the introduction, let us here briefly mention the
inflation scenario using the GUT Higgs boson, since it is an old topic
and was studied extensively in the past.  The graceful exit problem of
the original inflation relying on the first-order phase transition was
avoided in the new inflationary universe scenario (new inflation)
proposed by Linde~\cite{Linde:1981mu}. The phase transition in the new
inflation was of Coleman-Weinberg (CW) type~\cite{Coleman:1973jx},
where the inflaton was the GUT Higgs boson with the mass at the origin
being set to be zero. Although this scenario was very attractive, it
was soon realized that the CW correction arising from the gauge boson
loop makes the inflaton potential too steep to produce the density
perturbation of the correct magnitude, $\delta \rho/\rho \sim
10^{-5}$~\cite{Starobinsky:1982ee}. In fact, the required magnitude of
the gauge coupling constant was many orders of magnitude smaller than
the expected value of the unified gauge coupling constant, which
clearly implied that some modification was needed. One solution was to
consider a gauge singlet inflaton, which has extremely weak
interactions with the SM particles.  Although the inflation model may
lose its connection to the GUT in this case,\footnote{ We note that
  the SUSY GUT provides a natural framework for hybrid
  inflation~\cite{Copeland:1994vg}.  }  such gauge singlets are
ubiquitous in the string theory, and so, one of them may be
responsible for the inflation. There have been many works along this
line~\cite{Kachru:2003sx}.  Another way to resolve the problem is to
introduce SUSY. Then the CW potential becomes suppressed because of
the cancellation among bosonic and fermionic degrees of freedom
running the loop~\cite{Ellis:1982ed}.  In this paper we will explore
the latter possibility in detail and estimate the required size of the
SUSY breaking.

 \vspace{5mm}

The rest of the paper is organized as follows. In Sec.~\ref{sec:2} we
discuss the inflationary dynamics of the U(1)$_{\rm B-L}$ Higgs boson
considering only the tree-level contributions, anticipating that the
CW potential will be partially canceled by SUSY.  We will consider the
radiative corrections to the inflaton potential and derive the upper
bound on the SUSY breaking scale in Sec.~\ref{sec:3}.  We discuss
implications of our scenario in Sec.~\ref{sec:4}. The last section is
devoted for discussion and conclusions.

\section{Set-up and inflaton dynamics}
\label{sec:2}
In the following we use $\phi$ to denote the Higgs boson responsible
for the U(1)$_{\rm B-L}$ breaking. Here we do not assume SUSY to allow
a situation in which the inflation scale is lower than the SUSY
breaking scale. We will discuss the supersymmetric version in the next
section.

 Let us consider an inflaton potential given by
\beq
V \;=\; V_0 - m_0^2 |\phi|^2 - \lambda_n \frac{|\phi|^{2n}}{M_*^{2n-4}} + \lambda_m \frac{|\phi|^{2m}}{M_*^{2m-4}},
\label{infV}
\eeq
where $m$ and $n$ are integers satisfying $m > n \geq 2$ and $M_*$ is
a cut-off scale of the theory. We expect $M_*$ to be not far from the
GUT scale $\sim \GEV{15}$. The CW potential arising from the B$-$L
gauge boson and the right-handed neutrino loops will be considered in
the next section. For the moment we focus on the inflationary dynamics
using the tree-level contributions.

After the $\phi$ breaks the U(1)$_{\rm B-L}$ gauge symmetry, its phase
component is absorbed into the massive B$-$L gauge boson. So we focus on its
radial component:
\beq
\varphi \;\equiv\; \sqrt{2} |\phi|.
\eeq
In terms of $\varphi$, we can write the scalar potential as
\beq
V(\varphi)\;=\;V_0 - \frac{m_0^2}{2} \varphi^2 - \frac{\kappa}{2n} \frac{\varphi^{2n}}{M_*^{2n-4}}
			+ \frac{\lambda}{2m} \frac{\varphi^{2m}}{M_*^{2m-4}},      \label{Vinf}
\eeq
with
\bea
\kappa &\equiv & \frac{n \lambda_n}{2^{n-1}},\\
\lambda &\equiv & \frac{m \lambda_m}{2^{m-1}}.
\eea
This potential has a global minimum at $\varphi=\varphi_{\rm min}$ given by
\beq
\varphi_{\rm min} \;=\; \lrfp{\kappa}{\lambda}{\frac{1}{2(m-n)}} M_*,     \label{phimin}
\eeq
which gives the U(1)$_{\rm B-L}$ symmetry breaking scale at low
energy.  For the above effective theory description to be valid,
$\varphi_{\rm min} \lsim M_*$ must be satisfied.  Requiring that the
cosmological constant vanishes at the minimum, we obtain
\beq
V_0 \;=\; \lrf{m-n}{2mn} \lrfp{\kappa^m}{\l^n}{\frac{1}{m-n}} M_*^4.    \label{V0}
\eeq
Here we have assumed that the mass term is negligibly small compared
to the higher order terms at the potential minimum. 

The B$-$L breaking scale can be inferred from the neutrino oscillation
data as follows.  The Majorana mass $m_N$ for the right-handed
neutrino $\nu_R$ is related to the B$-$L breaking scale through the following
interaction,
\beq
{\cal L} \;=\; -\frac{y_N}{2} \phi \,{\bar {\nu}_R^c} \nu_R + {\rm h.c.}.
\label{phinur}
\eeq
and we obtain $m_N = y_N \varphi_{\rm min}/\sqrt{2}$. The coupling constant
$y_N $ is expected to be order unity for the heaviest $\nu_R$. Then the B$-$L
breaking scale $\varphi_{\rm min}$ is estimated to be about
$\GEV{15}$, close to the GUT scale, using the seesaw
formula~\cite{seesaw}.

For simplicity we drop the mass term, setting $m_0 = 0$, in the
following analysis.  All the results remain almost intact as long as
the mass is much smaller than the Hubble parameter during inflation.
Also, such a small mass may be 
favored since the total e-folding number of the inflation will be longer.

The inflation takes place if the initial position of $\varphi$ is
sufficiently close to the origin.  Since the inflaton has 
couplings to the B$-$L gauge boson, the right-handed neutrinos and the SM particles
(through the U(1)$_{\rm B-L}$ gauge interaction), we expect that
the initial position of $\varphi$ before the inflation starts is
naturally close to the origin, assuming the presence of thermal plasma
in the universe.\footnote{ We assume that the universe
  had experienced another inflation before the last inflation by
  $\phi$ started. The radiation may have come from the decay of the
  inflaton responsible for the preceding inflation. Even without the
  radiation, the initial position may be naturally set to be at the
  origin by e.g. the Hubble-induced mass since it is the enhanced
  symmetry point.  } The inflation ends at the point where the
slow-roll conditions are violated, namely, one of the slow-roll
parameters $\eta$ becomes order unity,
\beq
\eta \equiv M_p^2 \frac{V''}{V} \simeq -1,
\eeq
where $M_p \simeq 2.4 \times 10^{18}$\,GeV is the reduced Planck mass. 
This occurs at
\bea
\varphi_{\rm end} &=& \left[ 
\frac{m-n}{2mn} \frac{1}{2n-1} \lrfp{\k}{\l}{\frac{n}{m-n}} M_*^{2n} M_p^{-2} 
\right]^\frac{1}{2(n-1)},\non\\
&=& \left[ 
\frac{m-n}{2mn} \frac{1}{2n-1}
\right]^\frac{1}{2(n-1)} \varphi_{\rm min}^\frac{n}{n-1} M_p^{-\frac{1}{n-1}}.
\eea
The field value of $\varphi$ at $N$ e-foldings before the end of
inflation, $\varphi_N$, can be estimated as follows. The e-folding
number $N$ is given by
\bea
N &\simeq& \int_{\varphi_N}^{\varphi_{\rm end}} -\frac{3H^2}{V'} d\varphi
\simeq \frac{V_0 M_*^{2n-4}}{2(n-1)M_p^2 \kappa }\varphi_{N}^{2-2n},
\eea
therefore we obtain
\bea
\varphi_N \;\simeq\; \lrfp{1}{N}{\frac{1}{2(n-1)}}  \lrfp{2n-1}{2(n-1)}{\frac{1}{2(n-1)}} \varphi_{\rm end},
\eeq
where $\varphi_N \ll \varphi_{\rm end}$ is assumed.  The slow-roll
parameter $\eta$ at $\varphi=\varphi_N$ is given by
\beq
\eta \;\simeq\; - \frac{2n-1}{2(n-1)} \frac{1}{N}.
\eeq
The other slow-roll parameter, $\epsilon$, is much smaller than
$|\eta|$, which is typically the case in the new inflation model.
Thus the scalar spectral index, $n_s$, is then given by
\beq
1-n_s \;=\;  \frac{2n-1}{n-1} \frac{1}{N}.   \label{ns}
\eea
In the limit $n\gg 1$, it approaches to $n_s = 0.96$ for $N = 50$,
which is close to the center value of the WMAP
result~\cite{Komatsu:2010fb}. For $n = 2(3)$, the spectral index is
about $0.94(0.95)$, which is also consistent with observation. As we
will see below, $n \geq 3$ is needed for non-thermal leptogenesis to
work unless there is a degeneracy among the right-handed neutrino
masses.  Thus, the spectral index is predicted to be between $0.94$
and $0.96$ in the simple case where the inflaton potential is dominated
by a single monomial term during the last $50$ e-foldings.  The precise value of $n_s$ actually
depends on the details of the inflaton potential, and it is possible
to slightly modify the prediction.

In order to account for the density perturbation by the quantum fluctuation of the inflaton,
 we impose the WMAP normalization
condition~\cite{Komatsu:2010fb},
\beq
\Delta_{\cal R}^2 \;\simeq\; 2.42 \times 10^{-9},
\eeq
where $\Delta_{\cal R}$ denotes the power spectrum of the curvature
perturbation ${\cal R}$.  In terms of the inflaton potential, it is
given by
\beq
\frac{V_0^3}{M_p^6 V'^2} \;\simeq\; 2.9 \times 10^{-7}.
\eeq
Using \EQ{V0}, we obtain
\beq
\kappa \;\simeq\;  2.9 \times 10^{-7} \lrfp{1}{2(n-1) N}{\frac{2n-1}{n-1}}\lrfp{2mn}{m-n}{\frac{n-2}{n-1}}
\lrfp{\varphi_{\rm min}}{M_*}{-\frac{2n(n-2)}{n-1}} \lrfp{M_p}{M_*}{\frac{2(n-2)}{n-1}}.     \label{kappa}
\eeq

In addition to $n$ and $m$, the inflaton potential of our interest has
four parameters: $V_0$, $\kappa$, $\lambda$, and $M_*$.  $V_0$ is
determined by requiring the vanishing cosmological constant at the
true vacuum, see (\ref{V0}).  Instead of $\lambda$, we prefer to use
the physically relevant quantity $\varphi_{\rm min}$, which is the
B$-$L breaking scale, given by (\ref{phimin}).  The WMAP normalization
condition (\ref{kappa}) then fixes the value of $\kappa$.  As a
result, we can parametrize the potential by two parameters: the cutoff
scale $M_*$ and the B$-$L breaking scale $\varphi_{\rm min}$.

\begin{figure}[tbp]
\begin{center}
\includegraphics[scale=1.2]{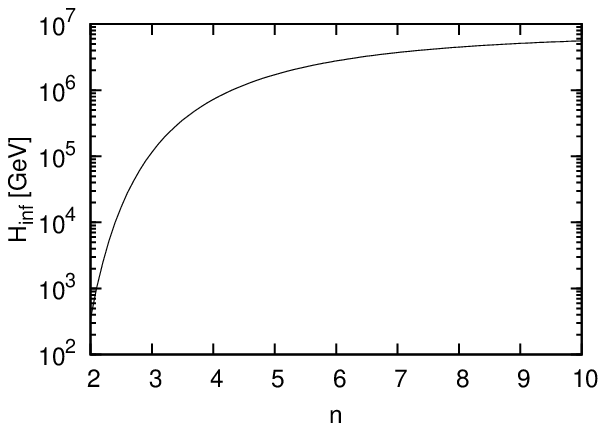}
\includegraphics[scale=1.2]{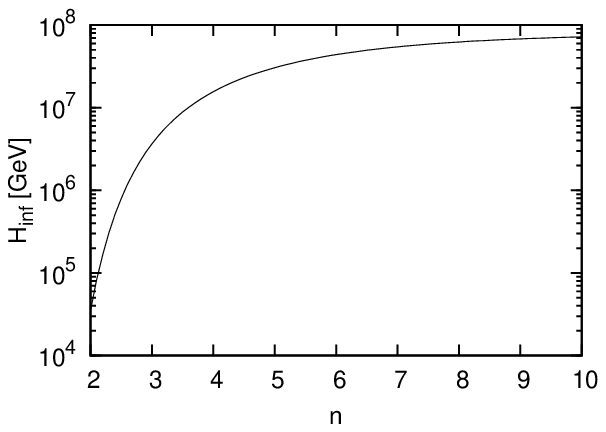}
\caption{
The Hubble parameter during inflation $H_{\rm inf}$ with respect to $n$. Here we set $m=n+1$.
The left is for $\varphi_{\rm min}=\GEV{14}$ and the right is for $\GEV{15}$.
The relation between $H_{\rm inf}$ and $n$ is almost same for $m=2n$.
	}
\label{fig:hinf}
\end{center}
\end{figure}

Since the inflaton potential is dominated by $V_0$ during
inflation, the Hubble parameter during inflation is estimated as
\beq
H_{\rm inf} \;=\;  3.1 \times 10^{-4} \lrfp{m-n}{2mn}{\frac{1}{2(n-1)}} 
\lrfp{1}{2(n-1) N}{\frac{2n-1}{2(n-1)}} \lrfp{\varphi_{\rm min}^n}{M_p}{\frac{1}{n-1}}.
\eeq
Note that the inflation scale is solely determined by the B$-$L
breaking scale, independent of $M_*$.  The inflation scale is shown in
Fig.~\ref{fig:hinf} for $\varphi_{\rm min}=\GEV{14}$ (left) and
$\GEV{15}$ (right).  One can see that, in the case of  $\varphi_{\rm min} = 10^{14}(10^{15})$\,GeV,
the Hubble parameter lies in the range of $10^2(10^4)$\,GeV and $10^7(10^8)$\,GeV.
%
In the case of $n=2(3)$, the Hubble parameter lies in the range of $10^2(10^5)$\,GeV
$\sim 10^4(10^6)$\,GeV.

\begin{figure}[tbp]
\begin{center}
\includegraphics[scale=1.2]{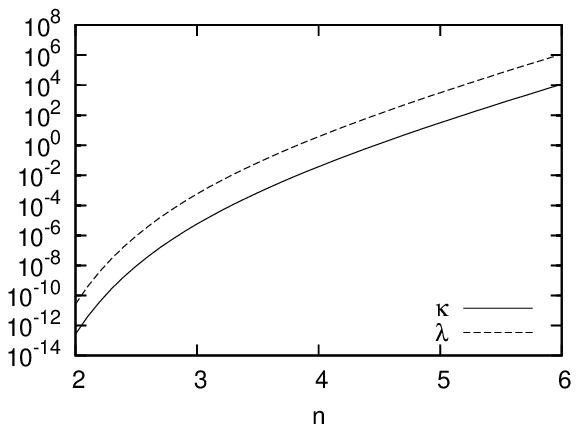}
\includegraphics[scale=1.2]{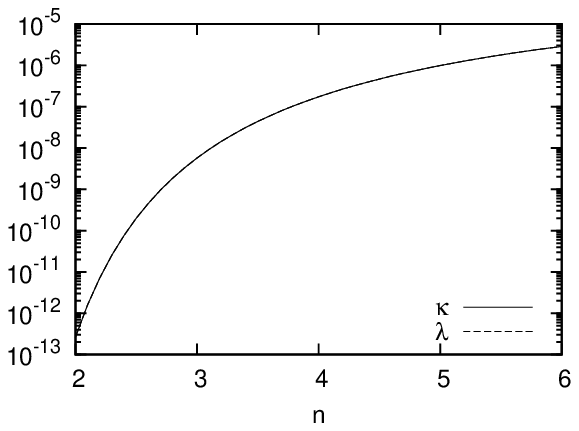}
\includegraphics[scale=1.2]{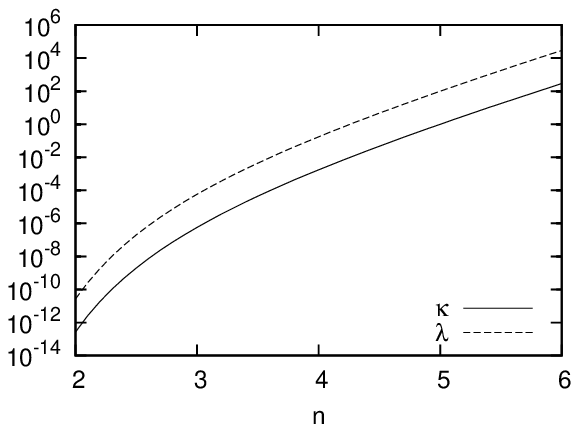}
\caption{
The $\kappa$ and $\lambda$ with respect to $n$. We set $m=n+1$.
 The top left is for $\varphi_{\rm min}= \GEV{14}$ and  $M_*= \GEV{15}$,
 the top right is for $\varphi_{\rm min}= \GEV{15} $ and  $M_*= \GEV{15}$,
 and the bottom is for $\varphi_{\rm min}= \GEV{15}$ and $M_*= \GEV{16}$.
 $n$ is bounded above as $n \leq 4$ for the top left and bottom cases. 
If $m=2n$, the bound becomes $n \leq 3$ for these cases. 
	}
\label{fig:kl}
\end{center}
\end{figure}

The values of $\kappa$ and $\lambda$ are also plotted in
Fig.~\ref{fig:kl} with $\varphi_{\rm min}= \GEV{14}$ and $M_*=
\GEV{15}$ (top left), $\varphi_{\rm min}= \GEV{15} $ and $M_*=
\GEV{15}$ (top right), and $\varphi_{\rm min}= \GEV{15}$ and $M_*=
\GEV{16}$ (bottom).  In the case of $\varphi_{\rm min} < M_*$, the
values of $n$ and $m$ are bounded above to avoid too large numerical
coefficients.  In the case of the top left and bottom panels, we
obtain $n \leq 4$.  However, this is sensitive to the relative
magnitude of $\varphi_{\rm min}$ and $M_*$. Indeed, for $\varphi_{\rm
  min} \approx M_*$, there is no such upper bound, and both $\kappa$
and $\lambda$ asymptote to $\sim 10^{-5}$ as $n$ becomes 
large (see the top right panel in Fig.~\ref{fig:kl}).

Thus, if we allow fine-tuning of the parameters, $m_0$, $\kappa$ and
$\lambda$, the successful inflation takes place using the B$-$L Higgs
boson. The inflation scale is typically very low, and only negligible
amount of the tensor mode is generated.  Note however that we have
only considered the tree-level potential. As we will see in the next
section, the radiative corrections generically spoil the inflationary
dynamics.

\section{Radiative correction to the inflaton potential}
\label{sec:3}

Now let us turn to the issue of the radiative correction to the
inflaton potential.  The inflaton, the B$-$L Higgs boson,
necessarily couples to the U(1)$_{\rm B-L}$ gauge boson.
Furthermore, it is expected to be coupled to the right-handed neutrinos to generate 
large Majorana masses. Due to these interactions,
 the inflaton potential receives corrections at the one-loop level. 
 The general form of the CW effective potential
is given by~\cite{Coleman:1973jx}
\begin{equation}
	V_{\rm CW} = \frac{1}{64\pi^2}\left[ 
		\sum_{B=\rm boson} m_B^4\left(\ln \frac{m_B^2}{\mu^2}-\frac{3}{2}\right)
		-\sum_{F=\rm fermion} 2m_F^4\left(\ln \frac{m_F^2}{\mu^2}-\frac{3}{2}\right)
	\right],    \label{CW}
\end{equation}
where the sum over bosons counts a real scalar, and that over fermions
counts a Weyl fermion.  Here the subscript $B$ denotes bosons, and it
includes the U(1)$_{\rm B-L}$ gauge boson, while $F$ denotes fermions
including the right-handed neutrinos. Since the masses of the U(1)$_{\rm B-L}$ gauge boson as
well as the right-handed neutrinos depend on the inflaton field $\varphi$,
the inflaton potential receives the CW correction.
For the moment we drop the contribution
from the right-handed neutrinos and focus on the gauge boson contribution.

In fact, it is well known that the CW potential arising from the gauge
boson loop makes the effective potential so steep that the resultant
density perturbation becomes much larger than the observed
one~\cite{Starobinsky:1982ee}.  One way to solve the problem is to
consider a gauge singlet inflaton. Here we stick to the inflation
model using the B$-$L Higgs boson and explore other
possibilities. Then we need to suppress the radiative correction
somehow.\footnote{ It is not possible to cancel the CW potential by
  tuning the tree-level potential, because of the logarithmic
  factor. Here we assume that the effective theory below $M_*$ is
  regular and that there are no additional light degrees of freedom
  other than SM+$\nu_R$+U(1)$_{\rm B-L}$$(+ {\rm SUSY})$.  }

One way to cancel or suppress the CW potential is to introduce SUSY.
In the exact SUSY limit, contributions from boson loops and fermion
loops are exactly canceled out.  However, if SUSY is broken, the
non-vanishing CW corrections remain.

In SUSY, two U(1)$_{\rm B-L}$ Higgs bosons are required  for anomaly
cancellation.  Let us denote the corresponding superfields as
$\Phi(+2)$ and ${\bar \Phi}(-2)$ where the number in the parenthesis
denotes their B$-$L charge.  The $D$-term potential vanishes along the
$D$-flat direction $\Phi {\bar \Phi}$, which is to be identified with
the inflaton. Actually, a linear combination of the lowest components
of $\Phi$ and $\bar{\Phi}$ corresponds to $\phi$.

The gauge boson as well as the scalar perpendicular to the $D$-flat
direction have mass of $m_B^2=g^2\varphi^2$. On the other hand, there
are additional fermionic degrees of freedom, U(1)$_{\rm B-L}$ gaugino
and the B$-$L higgsino, whose mass eigenvalues are given by $m_F =
g\varphi \pm \tilde m$, where $\tilde m$ denotes the SUSY breaking mass
for the B$-$L gaugino.
Because of the SUSY breaking mass $\tilde m$, the CW potential does not vanish and the
inflaton receives a non-zero correction to its potential.\footnote{
  The possibility that the gauge non-singlet inflaton is protected
  from radiative corrections by SUSY was pointed out in
  Ref.~\cite{Ellis:1982ed}, but the estimate of the CW potential
  during inflation is not correct and is different from ours by many
  orders of magnitude.  }
Inserting the field dependent masses into the CW potential (\ref{CW}), and expanding it by $\tilde m/ (g\varphi)$, we find
\begin{equation}
	V_{\rm CW}(\varphi) \simeq \frac{g^2}{8\pi^2} \left(1- 3\ln\frac{g^2\varphi^2}{\mu^2} \right) \tilde m^2 \varphi^2 .
	\label{CW2}
\end{equation}
Thus, in the presence of SUSY, the CW potential becomes partially
canceled and the dependence of the inflaton field has changed from
quartic to quadratic as long as ${\tilde m} \ll g \varphi$. Note that the correction still contains a
logarithmic factor, which is not negligible if we consider the whole
evolution of the inflaton from the origin.

If the mass of the CW potential exceeds the Hubble parameter, the inflaton does not slow-roll
and inflation does not occur. 
In order not to disturb the inflationary dynamics studied in the
previous section, therefore, the mass correction due to the CW potential should
be sufficiently small for a certain range of the inflaton field.   We
require that the curvature of the inflaton potential is much smaller
than the Hubble parameter everywhere from $\varphi = H_{\rm inf}/2\pi$
to $\varphi = \varphi_{N}$. This is a reasonable assumption because,
in the new inflation scenario, the universe likely experiences the
eternal inflation when the inflaton is near the origin where the
quantum fluctuation dominates the dynamics.  The presence of the
eternal inflation may be  favored in the landscape, since it can
compensate the required fine-tuning of the parameters.  For $n = 2 -
10$ with $N=50$, we estimate the logarithmic factor as
$\ln(\varphi_{\rm N}^2/(H_{\rm inf}/2 \pi)^2) \sim 30$.  Therefore,
the following inequality must be satisfied,
\beq
 \frac{g^2}{8\pi^2} 90\,  \tilde m^2 \;\lsim\; 0.01 H_{\rm inf}^2,
 \label{cond}
\eeq
therefore,
\begin{equation}
	\tilde m \; \lsim \;  0.1\,H_{\rm inf}.
	\label{bound}
\end{equation}
The reason why we put $0.01$ in the right-handed side of \EQ{cond} is
that the observed spectral index, $1-n_s$, is of ${\cal O}(0.01)$.\footnote{
Here we are interested in the inflation models which explain the observed data
in our universe, since it is hard to estimate the likelihood of universes with
observables taking different values.
}  Thus, 
the SUSY breaking scale should be one order of magnitude smaller than
the Hubble parameter during inflation, if one requires that the
successful inflation take place. This is the main result of this
paper.

In the gravity mediation, the soft SUSY breaking mass for B$-$L gaugino
is considered to be comparable to the soft SUSY masses for the SSM particles.
For simplicity we assume the gravity mediation in the following.
We will come back to this issue and consider other possibilities in Sec.~\ref{sec:4}.

We emphasize here that this novel bound on the soft SUSY
breaking mass is derived from the requirement that the inflation
should occur.  Even if high-scale SUSY breaking scale is favored in
the string landscape, the anthropic pressure by the inflation
constrains the SUSY breaking scale below the inflation scale.  Also,
in this case we have a prediction that the SUSY breaking scale should
be close to the inflation scale.  For the choice of $n=2$ and
$\phi_{\rm min}= 10^{15}$GeV, the inflation scale is given by $H_{\rm
  inf} ={\cal O}(10^4)$GeV. Thus, assuming that the soft SUSY mass is
close to the upper bound, we obtain ${\tilde m} \;=\; {\cal
  O}(1)$\,TeV.  That is to say, the SUSY breaking scale, $\tilde m$,
happens to be close to the weak scale, independently of the gauge
hierarchy problem. If this is the case, SUSY may be discovered in the
TeV range at the LHC. As $n$ becomes larger, the SUSY breaking scale
can be higher, but it is generically smaller than ${\cal
  O}(10^3)$\,TeV.    The upper bound may be saturated in the landscape
if there is bias toward the longer duration of the inflationary phase.
In this case,    the SUSY particles are unlikely to be discovered at the LHC unless the SSM mass spectrum
has some hierarchical structure, but we may be able to see the hint for
the SUSY breaking scale of ${\cal O}(10^3)$\,TeV from the large 
radiative correction to the SM Higgs boson mass, which may fall in the range
of $m_H \gtrsim 140$\,GeV or so~\cite{Wells:2004di}.

\vspace{3mm}

Since the SUSY remains a good symmetry below the inflation scale, it
is possible to write down the inflation model in SUSY.  Since the $\phi$
cannot have a large $F$-term when it is near the origin,  there
must be another superfield $S$ which has a non-vanishing $F$-term as in the
usual SUSY inflation models. The model is similar to the two-field new
inflation model in Ref.~\cite{Asaka:1999jb}.  The superpotential is
given by
\beq
W\;=\;  S \left(v^2 - k \frac{\phi^{2 \ell}}{M_*^{\ell -2}}\right),
\eeq
where $v$ determines the inflation scale and $k$ is a coupling
constant.  It is possible to make the inflaton mass sufficiently small
for a certain K\"ahler potential so that the inflation takes place. Since
$S$ is a gauge singlet field, it does not modify our argument
in the previous section. Note that the constant term in the superpotential,
$W_0 =m_{3/2} M_p^2$, does not affect the inflation dynamics in this model,
because $S$ is stabilized near the origin during and after inflation.
This should be contrasted to the single-field new inflation~\cite{Izawa:1996dv} or 
hybrid inflation~\cite{Copeland:1994vg} (see also Refs.~\cite{Nakayama:2010xf,Buchmuller:2000zm}).

\vspace{3mm}

Lastly we note that it is actually possible to cancel the CW potential
by tuning the coupling $y_N$, because the contributions from the
right-handed neutrinos are accompanied with the minus sign in
Eq.~(\ref{CW}). This may be an interesting possibility, but it is not
certain whether $y_N$ efficiently scans the desired range
independently of the gauge coupling in the landscape.  If this is the
case, however, the inflation model studied in the previous section
works, and most of the results concerning the inflationary dynamics
(except for SUSY) in this paper remain valid.

Barring cancellation, a similar bound for a SUSY breaking mass
for the right-handed sneutrino can be derived.  Assuming the coupling $y_N$ of order unity
for the heaviest right-handed neutrino, we have
\beq
{\tilde m}_N \;\lesssim\;0.1 H_{\rm inf},
\eeq
where the right-handed sneutrino mass squared is given by $m_N^2 + {\tilde m}_N^2$,
while the right-handed neutrino mass is $m_N$.
For a generic K\"ahler potential, ${\tilde m}_N$ is expected to be 
of order the gravitino mass $m_{3/2}$.

\section{Cosmological and phenomenological implications}
\label{sec:4}
Here we summarize features of our scenario and discuss its implications.

\vspace{3mm}

{\it The inflation model based on the minimal set of particles} \\ We
have built an inflation model with the inflaton being identified with
the Higgs boson responsible for the breaking of the U(1)$_{\rm B-L}$
gauge symmetry. The particle content in our set-up is minimal in some
sense: the SM particles, the right-handed neutrinos, U(1)$_{\rm
  B-L}$ gauge symmetry and its associated Higgs boson $\phi_{\rm
  B-L}$, and their superpartners at a certain energy scale.

\vspace{3mm}

{\it The SUSY breaking scale}\\ The CW potential spoils the inflaton
dynamics if the sot SUSY breaking mass of the B$-L$ gaugino is higher
than the Hubble scale during inflation.  This sets an upper bound on
the SUSY breaking scale as ${\tilde m},\, {\tilde m}_N < 0.1 H_{\rm inf}$.

So far we have not specified how the SUSY breaking in another sector
is transmitted to the U(1)$_{\rm B-L}$ gaugino. In the gravity
mediation, we expect ${\tilde m} \sim m_{3/2}$, and the soft SUSY
breaking masses $m_s$ and $m_{\lambda}$ for the SSM particles will be the same order.
Here $m_s$ and $m_\lambda$ collectively represent the soft SUSY breaking mass
for scalars and gauginos, respectively. 
In the anomaly mediation, ${\tilde m}$ as well as $m_\lambda$ could be
loop-suppressed with respect to the gravitino mass. In particular,
since ${\tilde m}_N$ is expected to be of order $m_{3/2}$ for a generic (non-sequestered) 
K\"ahler potential,   we obtain ${\tilde m}_N \sim m_{3/2} \gg {\tilde m},\, {\tilde m_\lambda}$.\footnote{
The B$-$L gaugino mass could be larger if U(1)$_{\rm B-L}$ sector
has Planck-suppressed couplings with the SUSY breaking sector,
which may be realized in the extra
dimensional framework. }
Such a split mass spectrum is an interesting possibility when there is a
bias toward high-scale SUSY.  We then expect
$m_{3/2} \sim 10^3$\,TeV, and the SSM gauginos are in the TeV
range.  In particular, the Wino LSP of mass $\sim 3$\,TeV would be a 
candidate for dark matter (DM)~\cite{Hisano:2006nn}.

In the gauge mediation, the SUSY breaking mass for U(1)$_{\rm B-L}$
gaugino will be suppressed as $\varphi$ becomes larger. If we require
that the inflaton slow-rolls from around the origin to $\varphi_{\rm
  end}$, the gravitino mass can be much lower than ${\tilde m}$.  In
this case, the SUSY mass spectrum is such that all the superpartners
of the SM particles are in the TeV or higher, while the gravitino is
much lighter and is the lightest SUSY particle (LSP). Therefore the
gravitino is a candidate for DM in this case.

\vspace{3mm}

{\it Moduli problem}\\ 
In the present scenario the SUSY breaking
scale is bounded by the Hubble scale during inflation.  If there is a
bias toward higher SUSY breaking scale, this bound may be saturated.
Then, depending on the mediation mechanism of the SUSY breaking, 
the gravitino mass $m_{3/2}$ can be comparable to or even slightly larger
than $H_{\rm inf}$.  If the modulus mass is of order $m_{3/2}$ or
heavier, the modulus abundance is suppressed during
inflation. Also the modulus mass is expected to be much heavier than the weak scale,
it decays anyway before the big bang nucleosynthesis. Thus,
the cosmological moduli problem can be solved, or at least relaxed considerably~\cite{Coughlan:1983ci}.  Moreover, it may avoid the modulus
destabilization during inflation~\cite{Kallosh:2004yh} in the KKLT
setup~\cite{Kachru:2003aw}, if $m_{3/2} \gtrsim H_{\rm inf}$ is (marginally) satisfied. 

\vspace{3mm}

\begin{figure}[tbp]
\begin{center}
\includegraphics[scale=1.2]{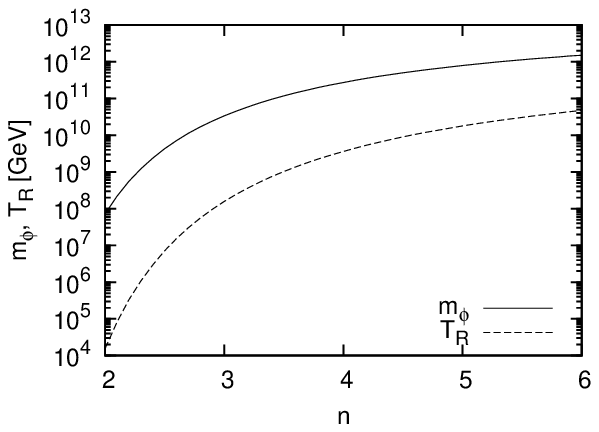}
\includegraphics[scale=1.2]{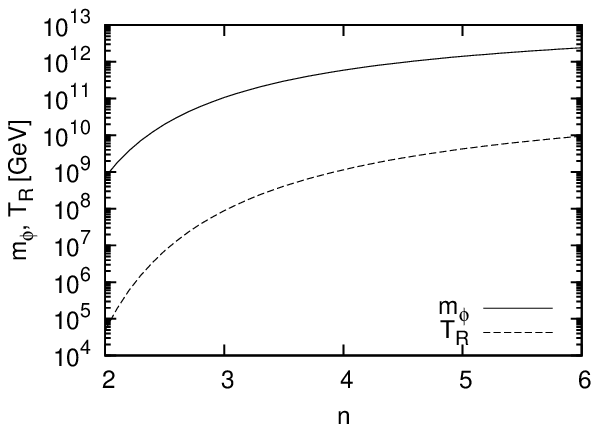}
\includegraphics[scale=1.2]{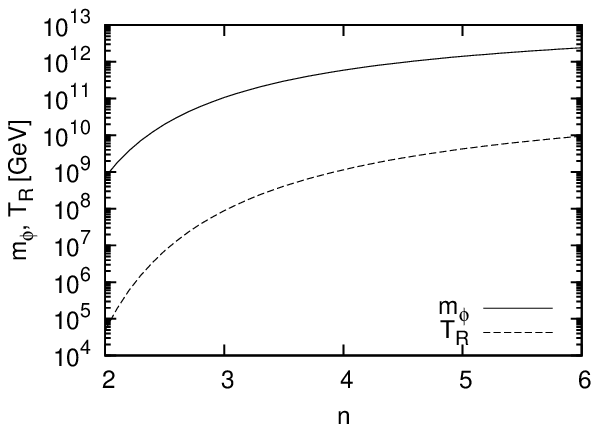}
\caption{
The B$-$L Higgs mass $m_\phi$ and the reheating temperature $T_R$ with respect to $n$, with $m=n+1$.
 The top left is for $\varphi_{\rm min}= \GEV{14}$ and  $M_*= \GEV{15}$,
 the top right is for $\varphi_{\rm min}= \GEV{15} $ and  $M_*= \GEV{15}$,
 and the bottom is for $\varphi_{\rm min}= \GEV{15}$ and $M_*= \GEV{16}$.
 We have set $m_N=0.1m_\phi$. In the case of $m=2n$, the relation between $T_R$ and $n$ 
is almost the same.
	}
\label{fig:TR}
\end{center}
\end{figure}

{\it Gravitino problem}\\ 
Since the inflaton mass is  light in our model, the reheating temperature is relatively low 
(see Fig.~\ref{fig:TR}). Note that the reheating temperature is correlated with the gravitino mass,
and that the cosmological bound on the gravitino is greatly relaxed as $m_{3/2}$ becomes large.
The overclosure bound on the LSP abundance produced from the gravitino decay can be avoided
if the R-parity is not conserved. 
Thus,
the gravitinos produced by thermal particle scatterings are cosmologically 
harmless. 
Furthermore, the non-thermal gravitino production from the inflaton decay~\cite{Kawasaki:2006gs}
is suppressed because the inflaton has a sizable coupling to the right-handed
neutrino (see (\ref{phinur})) and the branching fraction into the gravitinos is small.

\vspace{3mm}

{\it Baryogenesis }\\ The B$-$L Higgs couples to the right-handed
neutrino $\nu_R$ as \EQ{phinur}.    The B$-$L Higgs mass around the
potential minimum is given by
\begin{equation}
	m_\phi^2 = 2 \lambda (m-n) \frac{\varphi_{\rm min}^{2m-2}}{M^{2m-4}}.
\end{equation}
If $m_\phi > 2 m_N$, the Higgs can decay into a pair of the
right-handed neutrinos with the decay rate
\begin{equation}
	\Gamma_\phi \;\simeq\; \frac{y_N^2}{8\pi}m_\phi.
\end{equation}
The reheating temperature is then estimated to be $T_R = (10/\pi^2
g_*)^{1/4}\sqrt{\Gamma_\phi M_P}$ where $g_*$ counts the relativistic
degrees of freedom at $T=T_R$. We adopt in the following $g_*=228.75$.
In Fig.~\ref{fig:TR}, the inflaton mass and the reheating temperature
are plotted for $\varphi_{\rm min}= \GEV{14}$ and $M_*= \GEV{15}$ (top
left), $\varphi_{\rm min}= \GEV{15} $ and $M_*= \GEV{15}$ (top right),
and$\varphi_{\rm min}= \GEV{15}$ and $M_*= \GEV{16}$ (bottom).  Here
we have assumed that $m_N=0.1 m_\phi$ for simplicity and determined
$y_N$ accordingly.  It is seen that the reheating temperature exceeds
$10^6$GeV for $n \geq 3$.  Thus the baryogenesis through non-thermal
leptogenesis naturally works for $n\geq 3$~\cite{Asaka:1999yd}.  In
the case of $n=2$, the reheating temperature is about $10^4$\,GeV, and
we need to assume degeneracy among the right-handed neutrinos to
generate the right amount of the baryon asymmetry.

\vspace{3mm}

{\it Dark matter}\\ Since SSM particles are likely thermalized after
reheating, the LSP can be DM if the R-parity is conserved.  For
instance, as we have seen, the Wino can be DM in a certain situation.
However, there is an argument that the R-parity violation may be a
common phenomenon in the string landscape~\cite{Tatar:2006dc}. If so,
the dangerous operators leading to the proton decay must be absent due
to some other reason(s) and the lifetime of LSP in the SSM may be too
short to account for the DM.\footnote{It is possible that the lifetime
  of the Wino LSP is sufficiently long and its decay product
  contribute to the cosmic-ray spectrum~\cite{Shirai:2009fq}.}  Even in this case, 
  the gravitino LSP  may
serve as DM. The decay of the gravitino DM may provide an observable
signature in the the cosmic-ray spectrum~\cite{Ibarra:2007wg}.

There is also a well motivated DM candidate, the QCD
axion~\cite{Peccei:1977hh, Kim:1986ax}.  Since the inflation scale is
low, the Peccei-Quinn (PQ) symmetry is likely broken during inflation.
The magnitude of the axion isocurvature perturbation is estimated to
be $S_{\rm iso} \simeq H_{\rm inf} / (2 \pi f_a)$ where $f_a$ denotes
the PQ scale.  This is very small for typical values of $H_{\rm inf}$
and $f_a$, but if the inflation scale is on the high side, it is
marginally consistent with the current observation and it may be
detected by the Planck satellite.

The high quality of the PQ symmetry is often considered as a mystery,
since any global symmetries are expected to be explicitly broken by
Planck-suppressed operators according to the argument on the quantum
gravity~\cite{Banks:2010zn,Hellerman:2010fv}.  One explanation is that the QCD axion
arises from the string theory axions, and is subject to the other
moduli stabilization mechanism~\cite{Dine:2010cr}. There appears a
small number in the moduli stabilization, namely the ratio of the
gravitino mass to the Planck scale, which could be extremely small if
the SUSY persists to low-energy scales. This hierarchy, the gravitino
mass and the Planck scale may be responsible for the high-quality of
the PQ symmetry.  In our framework, therefore, the origin of the
high-quality of PQ symmetry could be a result of the inflationary
selection.

\vspace{3mm}

{\it The SM Higgs boson mass}\\ The SM Higgs boson mass weakly depends
on the soft masses of the SSM particles~\cite{Okada:1990vk,Ellis:1990nz,Haber:1990aw}.  In our
framework, the SUSY breaking scale can be as large as $\GEV{6}$, and
if it is on the high side, the Higgs boson mass will receive sizable
radiative corrections and becomes heavier than the case of the
weak-scale SUSY. The Higgs boson mass lies within the range of $125$\,GeV 
and $155$\,GeV~\cite{Wells:2004di}, which will be soon checked at the LHC.
Although the SUSY particles are beyond the reach of LHC
in this case, we may be able to obtain a hint for such a large SUSY breaking
from the SM Higgs boson mass.

\vspace{3mm}

{\it Spectral index and tensor mode}\\ As already calculated in
(\ref{ns}), the spectral index $n_s$ varies from $0.94$ to $0.96$
depending on $n$, although it could be increased or decreased by
further tuning the potential, which however is not needed.  The
amplitude of the tensor mode is negligibly small and it cannot be
detected by future observations.  Also no sizable non-Gaussianity is
generated.

\section{Discussion and Conclusions}
\label{sec:5}
In this paper we have built an inflation model with the inflaton being
identified with the Higgs boson responsible for the U(1)$_{\rm B-L}$
gauge symmetry breaking, in the minimal framework
SM+$\nu_R$+U(1)$_{\rm B-L}$.  Our main conclusion is that the soft
SUSY breaking masses for the SSM particles should be one order of
magnitude smaller than the Hubble parameter during inflation.  As a
result, the SUSY breaking mass is bounded above as, $m_s \lesssim
1{\rm\,TeV} - \TEV{3}$. It is intriguing that, requiring the inflation
model based on the B$-$L Higgs boson to work, the SUSY breaking scale
is bounded above and it happens to be close to the weak scale, without
relying on the conventional naturalness argument about the gauge
hierarchy problem. There are several implications of our
finding. First, the universes with the low-scale SUSY may be selected
by the inflationary dynamics. Even if there is a bias toward
high-scale SUSY in the string landscape, there is a hope that SUSY may
be found in the collider experiments such as LHC.  Also, the SUSY particles are
not very far from the weak scale, even when they are beyond the reach of LHC.
It should be
emphasized here that this conclusion is derived not relying on the
naturalness argument; the low-scale SUSY could emerge as a result of
the inflationary selection.  Second, even if SUSY is found at the LHC,
the fine-tuning issue of obtaining the correct electroweak breaking
(the little hierarchy problem) may not be a serious problem any more,
because the driving force for the low-scale SUSY is not the
fine-tuning issue, but the inflationary dynamics. Thus, the apparent
fine-tuning could be a result of combination of the low-scale matter
inflation and a bias toward high-scale SUSY.

Our inflation model has interesting cosmological implications. First,
since the inflaton is charged under the U(1)$_{\rm B-L}$ symmetry, it
is reasonable to expect that the inflaton has a sizable coupling with
the right-handed neutrinos, making the non-thermal leptogenesis
scenario attractive and viable.  Secondly, the inflaton may naturally
sit at the origin before the inflation starts because of its gauge
interactions with the high-temperature plasma. As the universe cools
down, the energy density of the plasma decreases, and the inflation is
considered to take place. Thirdly, since the inflationary scale is
much lower than the GUT scale, the size of the tensor fluctuation is
prohibitively small.  The scalar spectral index is expected to be in
the range of $n_s = 0.94 \sim 0.96$, but the precise value depends on
the detailed structure of the inflaton potential. The prime dark
matter candidate is the QCD axion. Since the inflation scale is low,
the isocurvature perturbation constraint on the QCD axion is not
stringent; if the inflation scale is on the high side, it is
marginally consistent and it may be detected by the Planck satellite.

We note that no topological defects are formed in the present model,
since the gauge symmetry is already broken during inflation.  This is
contrasted to the case of GUT hybrid inflation model, in which the
formation of topological defects is
inevitable~\cite{Jeannerot:2003qv}.

Since the soft SUSY breaking masses should be in the range between
$1$\,TeV and $10^3$\,TeV, the gauge coupling unification is improved
compared to the case without low-scale SUSY. In fact, the unification
looks reasonably good if $m_s \sim 10^3$\,TeV and $m_\lambda \sim $\,TeV~\cite{Wells:2004di}.
Again, this is due to the
inflationary selection.

Although we have focused on the B$-$L Higgs boson as the inflaton, it
is straightforward to apply our mechanism to the other GUT Higgs
bosons, and we expect similar conclusions about the SUSY breaking
scale can be reached. Detailed discussion on this issue is left for
future work.

\section*{Acknowledgment}
FT thanks M. Kawasaki for useful communication.  This work was
supported by the Grant-in-Aid for Scientific Research on Innovative
Areas (No. 21111006) [KN and FT], Scientific Research (A)
(No. 22244030 [KN and FT] and No.21244033 [FT]), and JSPS Grant-in-Aid
for Young Scientists (B) (No. 21740160) [FT].  This work was also
supported by World Premier International Center Initiative (WPI
Program), MEXT, Japan.



\end{document}